\documentclass[aps,pra,reprint,groupedaddress,amsmath,amssymb]{revtex4-1}
\pdfoutput=1

\usepackage{hyperref}
\DeclareMathOperator{\tr}{tr}
\newcommand{\ket}[1]{\ensuremath{\lvert #1 \rangle}}
\newcommand{\outprod}[2]{\ensuremath{\lvert #1 \rangle \langle #2 \rvert}}
\newcommand{\eqcl}[1]{\lbrack #1 \rbrack}

\begin{document}

\title{Complementarity has empirically relevant consequences for the definition of quantum states}

\author{Bradley A. Foreman}
\affiliation{Department of Physics, The Hong Kong University of Science and 
Technology, Clear Water Bay, Kowloon, Hong Kong, China}

%\date{\today}

\begin{abstract}
The Copenhagen interpretation of quantum mechanics, which first took shape in
Bohr's landmark 1928 paper on complementarity, remains an enigma. Although many
physicists are skeptical about the necessity of Bohr's philosophical
conclusions, his pragmatic message about the importance of the whole
experimental arrangement is widely accepted. It is, however, generally also
agreed that the Copenhagen interpretation has no direct consequences for the
mathematical structure of quantum mechanics. Here I show that the application of
Bohr's main concepts of complementarity to the subsystems of a closed system
requires a change in the definition of the quantum state. The appropriate
definition is as an equivalence class similar to that used by von Neumann to
describe macroscopic subsystems. He showed that such equivalence classes are
necessary in order to maximize information entropy and achieve agreement with
experimental entropy. However, the significance of these results for the quantum
theory of measurement has been overlooked. Current formulations of measurement
theory are therefore manifestly in conflict with experiment. This conflict is
resolved by the definition of the quantum state proposed here.
\end{abstract}

\maketitle

\paragraph*{Introduction.}

Despite nearly a century of effort, research into the foundations of quantum
mechanics remains a Tower of Babel. Weinberg describes the situation well
\cite{[] [{; reprinted in }] Weinberg2017a, *Weinberg2018}: ``It is a bad sign
that those physicists today who are most comfortable with quantum mechanics do
not agree with one another about what it all means.'' Many physicists believe
that the most contentious issue, the measurement problem, was solved long ago by
some variant of orthodoxy \cite{Englert2013}, and that the only remaining
problem is with ``a set of people'' \cite{[{Attributed to H. Mabuchi by }] [{,
p.\ 166.}] Fuchs2011}. However, the orthodox solution has been criticized by
many others, including Gell-Mann, who lamented that ``Niels Bohr brainwashed a
whole generation of theorists into thinking that the job was done [in the
1920s]'' \cite{GellMann1979}.

Some of the more prevalent tongues heard in the Tower are those of the
Copenhagen \cite{Bohr1928,Bohr1935,Bohr1949} and Princeton
\cite{vonNeumann1955,Wigner1967} schools (each of which has some claim to
orthodoxy), the Everett relative-states or many-worlds theory
\cite{Everett1957,Wheeler1957,Saunders2010}, consistent histories
\cite{Laloe2019}, pilot-wave or hidden-variables models \cite{Laloe2019},
dynamical-reduction models \cite{BassiLochan2013,ArndtHornberger2014}, and a
range of approaches influenced by the concepts of quantum information, including
reconstruction efforts \cite{Ball2013} and QBism \cite{FuchsMerminSchack2014}. A
cross-section of the diversity of current opinion can be found in conference
surveys \cite{SchlosshauerKoflerZeilinger2013} and interviews
\cite{Schlosshauer2011}. Decoherence theory
\cite{JoosZeh2003,Zurek1991,Zurek2003,Schlosshauer2007} has become increasingly
popular, in no small part because it is said to have a foot in both the
Copenhagen and Everett camps \cite{Zurek1991,Zurek2003,Camilleri2009b}; with the
advent of quantum Darwinism \cite{Zurek2009}, decoherence theory is now also
viewed as a branch of quantum information theory.

The reconciliation of Bohr and Everett achieved thus far by decoherence theory
is, however, rather limited, with Copenhagen ideas entering mainly via the
language by which the results of decoherence theory can be described. The
difficulty that many physicists have experienced in putting the amorphous
Copenhagen philosophy to practical use is crystallized in the words of Mehra, as
reported by Bell \cite{BassiGhirardi2007}: ``Though Dirac appreciated and
admired Bohr greatly, he told me that he did not find any great significance in
Bohr's principles of correspondence and complementarity because they did not
lead to any mathematical equations.'' Bell responded \cite{BassiGhirardi2007} by
saying that ``I absolutely endorse this opinion'' and added that he could not
explain complementarity because he ``never got the hang of it.'' This opinion
seems to be shared by most authors of quantum-mechanics textbooks, whose work is
dominated by the mathematics of the Princeton school, with at most a few
paragraphs of verbal garnish on the Copenhagen philosophy.

%\cite{[{Similar but not identical statements are given on pp.\ 40--41 and 53--54
%of }] [{, and on pp.\ 330--331 of }] Mehra1972b, *MehraRechenberg2000}

Here I show that a unification of the core concepts of Bohr and Everett does
indeed lead to new equations. The key concepts in this regard are Bohr's
emphasis on the whole experimental arrangement---which implies that the
properties of a measured microscopic subsystem acquire meaning only through its
interaction with a macroscopic measuring apparatus and that ``unperformed
experiments have no results'' \cite{Peres1978}---and Everett's insistence that
all relevant subsystems, including those associated with a measuring apparatus
or observer, be included in the quantum-mechanical description. A merger of this
type can be viewed as a long-delayed but inevitable consequence of Bohr's
capitulation to Einstein by including the apparatus in the domain of
applicability of Heisenberg's uncertainty relations \cite{Bohr1949}. Such a
merger was much desired by Wheeler during his unsuccessful attempt to mediate
between Bohr and Everett \cite{Wheeler1957,Osnaghi2009}.

As shown here, a consistent combination of these ideas requires a change in the
definition of a quantum state. The appropriate definition is not as a ray or
density operator in Hilbert space, but as an equivalence class of stable
information. Vital components of this definition have appeared previously in the
work of von Neumann \cite{vonNeumann1955secV4}, Landau and Lifshitz
\cite{LandauLifshitz1977_sec7}, and Zurek \cite{Zurek2000}, but always in
isolation, never as a coherent whole. This formulation of the quantum state has
important experimental implications for the entropy of macroscopic subsystems
\cite{vonNeumann1955secV4}. A separate paper \cite{Foreman2019a} provides
further details of this theory and shows that this definition of a quantum state
solves---in the limited sense of emergence---the problem of outcomes, which is
the last outstanding part of the quantum measurement problem
\cite{Schlosshauer2007}. Perhaps this empirically meaningful unification of two
of the major strands of thought in the foundations of quantum mechanics may help
to resolve some differences of opinion in this field.

\paragraph*{Direct and indirect reductions.}

Let us start by considering a set $A$ of subsystems, each of which interacts
strongly with its environment and therefore decoheres rapidly. Such a subsystem
is said to be \emph{integrated} with its environment. The set of integrated
subsystems contains those exhibiting quasiclassical behavior, including
macroscopic measuring apparatuses. Subsystems in set $B$ may be \emph{isolated}
in the sense that they interact weakly with their environments, apart perhaps
from occasional strong interactions with measuring apparatuses in set $A$.
Integrated subsystems are typically macroscopic, whereas isolated subsystems are
typically microscopic. The composite system $AB$ is assumed to be
\emph{closed}---i.e., it does not interact with anything else.

In this closed system, a state vector generally has the entangled form
\begin{equation}
\ket{\psi} = \sum_{\alpha, \beta} c_{\alpha \beta} \ket{\psi^{A}_{\alpha}}
\ket{\psi^{B}_{\beta}} \quad (c_{\alpha \beta} \in \mathbb{C}) ,
\label{eq:psi_general}
\end{equation}
in which $\{ \ket{\psi^{A}_{\alpha}} \}$ and $\{ \ket{\psi^{B}_{\beta}} \}$ are
complete orthonormal basis sets in $A$ and $B$, respectively. This can always be
written as a relative-state expansion \cite{LandauLifshitz1977_sec7,Everett1957}
\begin{equation}
\ket{\psi} = \sum_{\alpha}  \ket{\psi^{A}_{\alpha}} \ket{\varphi^{B}_{\alpha}} ,
\label{eq:psi_relative}
\end{equation}
in which
\begin{equation}
\ket{\varphi^{B}_{\alpha}} = \sum_{\beta} c_{\alpha \beta}
\ket{\psi^{B}_{\beta}} \label{eq:phi_relative}
\end{equation}
is the state in $B$ relative to $\ket{\psi^{A}_{\alpha}}$ in $A$.

Consider now a typical quantum measurement situation, whereby a macroscopic
apparatus in $A$ interacts with a microscopic subsystem in $B$. The outcome of
such a measurement is conventionally described as a reduction process generated
by an exhaustive orthonormal set $\mathcal{P}$ of projection operators $P_i$:
\begin{equation}
\mathcal{P} = \{ P_i \mid P_i^{\dagger} P_j = \delta_{ij} P_i, \sum_i P_i = 1 \}
. \label{eq:projector_set}
\end{equation}
Note that in decoherence theory, the set $\mathcal{P}$ and the subsystems used
to define $A$ and $B$ are not given \emph{a priori}. Rather, they are derived
from a variational principle known as the ``predictability sieve''
\cite{Zurek2003}, which selects projectors and subsystems \cite{Foreman2019a} by
the criteria of stability and predictability.

According to von Neumann \cite{vonNeumann1955} and most quantum-mechanics
textbooks, the projectors $P_i$ act either on the ``measured'' subsystem 
in $B$ or on all subsystems together.  However, as stressed by Landau and Lifshitz \cite{LandauLifshitz1977_sec7}, the measurement process is 
better described by using projectors that act nontrivially only in $A$:
\begin{equation}
P_i = P_i^{A} \otimes 1_{B} , \qquad P_i^{A} = \sum_{\alpha \in i}
\outprod{\psi^{A}_{\alpha}}{\psi^{A}_{\alpha}} . \label{eq:projector_A}
\end{equation}
The reason for this is simply that we obtain our information about the outcome
of the measurement from the measuring apparatus, not from the microscopic
subsystem. The vector subspace defined by the projector $P_i^{A}$ is usually
associated with a set of collective variables, such as those in which the center
of mass of the apparatus pointer lies in some given volume in coordinate space.
But in a typical quantum measurement situation, all of the states
$\ket{\psi^{A}_{\alpha}}$ in this subspace may have the same relative state
$\ket{\varphi^{B}_{\alpha}}$. The \emph{direct} reduction of $A$ generated by
$P_i$ thus leads to an \emph{indirect} reduction of $B$, in which the state of
$B$ is (in this example) reduced to the pure state $\ket{\varphi^{B}_{\alpha}}$.
This indirect reduction mechanism \cite{LandauLifshitz1977_sec7} is an explicit
implementation of Bohr's principle \cite{Bohr1928} that the properties of a
microscopic subsystem acquire meaning only through their correlations with those
of a quasiclassical apparatus.

The reduction process is most convenient to write in terms of density
operators $\rho$, where $\rho = \outprod{\psi}{\psi}$ in the special case of a
pure state described above. It can be separated conceptually into two stages,
the first of which is the elimination of interference between the alternatives
$i$:
\begin{equation}
\rho \to \hat{\rho} = \sum_{i} P_i \rho P_i . \label{eq:Lueders_stage_1}
\end{equation}
The second stage is the selection of one individual outcome from
among these alternatives:
\begin{equation}
\hat{\rho} \to \rho_i = \frac{P_i \rho P_i}{w_i} , \qquad w_i = \tr (\rho P_i) ,
\label{eq:Lueders_stage_2}
\end{equation}
in which $w_i$ is the probability of obtaining outcome $\rho_i$. This is the
L\"uders reduction process \cite{Lueders1950}, which is used in most textbooks
for the description of ideal measurements. Regardless of whether one chooses to
portray reduction as real, illusory, emergent, or merely the acquisition of
information, it must be taken into account if one is to make contact with
experimental physics \cite{Everett1957,JoosZeh2003}.

\paragraph*{Superselection rules and equivalence classes.}

The elimination of interference in Eq.\ (\ref{eq:Lueders_stage_1}) is commonly
described in decoherence theory as the emergence of an ``environment-induced
superselection rule'' \cite{Zurek1991,Zurek2003}. A brief detour
into the theory of superselection rules in algebraic quantum mechanics
\cite{Beltrametti1981} is helpful in explaining the reason for this.
Superselection rules are used to implement the hypothesis that, contrary to the
tacit assumption used in most elementary textbooks, not all operators in Hilbert
space correspond to observable quantities.

Let $\mathcal{S}$ be some set of operators in Hilbert space. Its commutant
$\mathcal{S}'$ is defined as the set of bounded operators that commute with all
operators in $\mathcal{S}$; the bicommutant is defined likewise as
$\mathcal{S}'' = (\mathcal{S}')'$. The \emph{algebra of observables} generated
by the set $\mathcal{S}$ is then defined to be $\mathcal{A} = \mathcal{S}''$; it
satisfies $\mathcal{A} = \mathcal{A}''$. The basic hypothesis under
consideration is that physical quantities in quantum mechanics are limited to
operators in $\mathcal{A}$, or operators whose spectral decomposition contains
only projectors in $\mathcal{A}$.

The nontrivial elements of $\mathcal{A}'$ (i.e., those not proportional to the
identity operator) are called \emph{superselection operators}. The \emph{center}
of $\mathcal{A}$ is defined as $\mathcal{Z} = \mathcal{A} \cap \mathcal{A}'$;
operators in $\mathcal{Z}$ are called \emph{classical observables}. In quantum
mechanics, it happens to be true that $\mathcal{A}' \subseteq \mathcal{A}$
\cite{Beltrametti1981}, which implies that $\mathcal{Z} = \mathcal{A}'$. In a
system with \emph{discrete} superselection rules (the only type considered
here), every superselection operator is of the form \cite{Beltrametti1981}
\begin{equation}
\Lambda = \sum_{i} \lambda_i P_i \qquad (\lambda_i \in \mathbb{C}, \ P_i \in
\mathcal{P}) .
\end{equation}
Here the set $\mathcal{P}$ is as shown in Eq.\ (\ref{eq:projector_set}),
although in general $P_i$ need not have the form given in Eq.\
(\ref{eq:projector_A}). In such a system, the choice of $\mathcal{P}$ completely
defines the algebra of observables, because $\mathcal{A}$ is generated by
$\mathcal{P}'$. Conversely, $\mathcal{P}$ is the set of nontrivial projectors
contained in $\mathcal{A}'$.

A \emph{state} in quantum mechanics is defined as a probability measure over the
set of all projectors $P \in \mathcal{A}$ \cite{Beltrametti1981}. For a system
with no superselection rules, it is well known that this probability measure can
be written as $\tr (P \rho)$, where $\rho$ is a density operator that belongs to
$\mathcal{A}$ \cite{Gleason1957}.  Quantum states are then in one-to-one
correspondence with density operators \cite{Gleason1957}.

However, this one-to-one correspondence is broken in a system with
superselection rules \cite{Beltrametti1981}. A quantum state can then only be
identified with an \emph{equivalence class} of density operators
\begin{equation}
\eqcl{\rho} = \{ \sigma \mid \sigma \sim \rho \}  , \label{eq:equivalence_class}
\end{equation}
in which the equivalence relation $\sigma \sim \rho$ is defined by
\begin{equation}
\sigma \sim \rho \ \Leftrightarrow \ \tr (P \sigma) = \tr (P \rho) \ \forall P
\in \mathcal{A} . \label{eq:equivalence_relation_standard}
\end{equation}
In such a system, a density operator $\rho$ is generally \emph{not} a member of
$\mathcal{A}$. However, the reduced density operator $\hat{\rho}$ in Eq.\
(\ref{eq:Lueders_stage_1}) belongs to both $\mathcal{A}$ and $\eqcl{\rho}$. In
fact, $\hat{\rho}$ is the \emph{unique} member of the set $\eqcl{\rho} \cap
\mathcal{A}$ \cite{Beltrametti1981}. The one-to-one correspondence between
$\eqcl{\rho}$ and $\hat{\rho}$ then implies that the equivalence class
$\eqcl{\rho}$ can be represented mathematically by the reduced density operator
$\hat{\rho}$. The equality $\rho = \hat{\rho}$ holds if and only if $\rho$
commutes with all $P_i \in \mathcal{P}$ \cite{Beltrametti1981}.

The fact that $\hat{\rho}$ is the canonical representative of the quantum state
$\eqcl{\rho}$ is the foundation for Jauch's theory of measurement
\cite{Jauch1964}. From this perspective, the first stage
(\ref{eq:Lueders_stage_1}) of the L\"uders reduction process is no change at
all, because $\rho$ and $\hat{\rho}$ correspond to the same quantum state
$\eqcl{\rho} = \eqcl{\hat{\rho}}$. The second stage (\ref{eq:Lueders_stage_2})
can then be viewed as merely the selection of one alternative $\rho_i$ from a
classical statistical mixture $\hat{\rho} = \sum_i w_i \rho_i$. The measurement
problem in quantum mechanics thus ``dissolves into a pseudoproblem''
\cite{Jauch1964}.

\paragraph*{Equivalence classes from complementarity.}

A major conceptual flaw in this approach is that, for a closed system, the
``observables'' in $\mathcal{A}$ have little to do with what is observed in an
experiment. According to this theory, both $\hat{\rho}$ and $\rho_i$ are
considered to be observables, because they belong to $\mathcal{A}$. However, in
an experiment, one certainly never has access to the full details of the density
operator $\rho_i$ at the time of reduction. All that one can really deduce from
the location of the apparatus pointer is that $\rho_i$ lies somewhere in the
subspace $\mathcal{M}_i$ defined by the projector $P_i$ in Eq.\
(\ref{eq:projector_A}):
\begin{equation}
\mathcal{M}_i = \{ \sigma \mid P_i \sigma = \sigma P_i = \sigma \} .
\label{eq:manifold}
\end{equation}
A more appropriate definition of the quantum state would take into account this
limit on the information that is actually accessible in an experiment.

Von Neumann has proposed an alternative definition of equivalence classes, for
the special case of systems consisting entirely of \emph{integrated} subsystems,
that is based directly on the meaningful information content of the reduced
state \cite{vonNeumann1955secV4}. Here von Neumann's definition is modified to
include the indirect reduction of isolated subsystems as well. In this
modified definition, the equivalence relation
(\ref{eq:equivalence_relation_standard}) is replaced with
\begin{equation}
\sigma \sim \rho \ \Leftrightarrow \ 
\tr (P_i Q \sigma) = \tr (P_i Q \rho) \ \forall P_i \in \mathcal{P} , Q \in
\mathcal{Q} , \label{eq:equivalence_relation_modified}
\end{equation}
in which $\mathcal{P}$ is defined in Eqs.\ (\ref{eq:projector_set}) and
(\ref{eq:projector_A}) and $\mathcal{Q}$ is the set of all projectors $Q = 1_{A}
\otimes Q_B$ that act nontrivially only in $B$. The fact that $Q_B$ is arbitrary
means that the isolated subsystems in $B$ are treated in the same way as a
system without superselection rules in the standard theory of Eq.\
(\ref{eq:equivalence_relation_standard}). However, $P_i$ is limited to the set
$\mathcal{P}$ used to perform the reduction that actually takes place at the
given time \cite{vonNeumann1955secV4}. This is a concrete implementation of
Bohr's principle of complementarity \cite{Bohr1928,Bohr1935,Bohr1949},
according to which (in the memorable phrase coined by Peres \cite{Peres1978})
``unperformed experiments have no results.''

The equivalence relation (\ref{eq:equivalence_relation_modified})
can be rewritten in the simpler form
\begin{equation}
\sigma \sim \rho \ \Leftrightarrow \ \tr_{A} (P_i \sigma) = \tr_{A} (P_i \rho) \
\forall P_i \in \mathcal{P} , \label{eq:equivalence_relation_simple}
\end{equation}
in which $\tr_A X = X_B$ denotes a partial trace over the subsystems in $A$, the
result of which is an operator in $B$. If the equivalence class
(\ref{eq:equivalence_class}) is now redefined in terms of this equivalence
relation, the reduced density operator $\hat{\rho}$ in Eq.\
(\ref{eq:Lueders_stage_1}) can no longer serve as the canonical representative
of $\eqcl{\rho}$. It must likewise be redefined as
\begin{subequations}
\label{eq:rho_hat_new}
\begin{gather}
\hat{\rho} = \sum_i w_i \rho_i , \qquad \rho_i = \rho_i^A \otimes \rho_i^B , \\
\rho_i^A = \frac{P_i^A}{d_i^A}  , \qquad
\rho_i^B =  \frac{\tr_A (P_i \rho)}{w_i} \quad (w_i \ne 0) , \label{eq:rho_i_A_B}
\end{gather}
\end{subequations}
in which $w_i$ is the same as before and $d_i^A = \tr P_i^A$. Here $\rho_i^A$
and $\rho_i^B$ are normalized density operators; $\rho_i^B$ is called the
\emph{conditional state} of $B$ given the state $\rho_i^A$ of $A$
\cite{SchumacherWestmoreland2010_p433}. Such conditional states are used in the
definition of quantum discord \cite{Zurek2000,OllivierZurek2001}, where they
play a role similar to that of the relative state in Eqs.\
(\ref{eq:psi_relative}) and (\ref{eq:phi_relative}). Once again, $\rho_i^B$
represents an \emph{indirect} reduction of $B$ that accompanies the direct
reduction of $A$ generated by $P_i$.

Equation (\ref{eq:rho_hat_new}) is the main result of this paper. This
modification of the reduced density operator $\hat{\rho}$ brings it back into
one-to-one correspondence with the modified equivalence class $\eqcl{\rho}$
(i.e., $\eqcl{\rho} = \eqcl{\sigma}$ if and only if $\hat{\rho} =
\hat{\sigma}$). This is a straightforward consequence of the orthogonality of
the projectors in the set (\ref{eq:projector_set}). Zurek has argued that such
orthogonality is necessary in order for the states of macroscopic subsystems to
be repeatedly accessible (a criterion closely related to the stability and
predictability criteria discussed below), using concepts very similar to the
equivalence classes of von Neumann \cite{Zurek2013}.

The entropy of $\eqcl{\rho}$ can also be defined as that of $\hat{\rho}$
\cite{vonNeumann1955secV4}, because $\hat{\rho}$ has the greatest von Neumann
entropy $S (\rho) = - k \tr (\rho \ln \rho)$ of any member of $\eqcl{\rho}$.
This can be seen directly from the definition (\ref{eq:rho_hat_new}), because
$\rho_i^A$ clearly has the maximum entropy of any state in the subspace
$\mathcal{M}_i^A$ defined by $P_i^A$, whereas the value of $\rho_i^B$ is fixed
by the equivalence relation (\ref{eq:equivalence_relation_simple}). The change
of reduced density operator $\hat{\rho}$ in going from Eq.\
(\ref{eq:Lueders_stage_1}) to Eq.\ (\ref{eq:rho_hat_new}) can thus be thought of
as a further stage of coarse-graining in which all information about the details
of any particular state in $\mathcal{M}_i^A$ is discarded.

\paragraph*{Experimental significance.}

This additional loss of information has crucial experimental consequences for
the entropy of macroscopic subsystems. As stressed by Jaynes
\cite{Jaynes1963,Jaynes1965,Jaynes1985}, $S (\rho)$ for any density operator
$\rho$ that satisfies a given set of macroscopic constraints (such as those that
define the collective variables of integrated subsystems) is related to the
experimental entropy $S_{\textrm{e}}$ measured under the same constraints by
$S(\rho) \le S_{\textrm{e}}$, where the equality holds if and only if $S (\rho)$
has been maximized with respect to all unconstrained variables in $\rho$. But
this means that $S(\hat{\rho}) = S_{\textrm{e}}$ for the reduced state
(\ref{eq:rho_hat_new}), whereas $S(\hat{\rho}) \ne S_{\textrm{e}}$ for the
L\"uders reduced state (\ref{eq:Lueders_stage_1}), because the latter does not
satisfy the maximum-entropy condition stated in the previous paragraph.
Therefore, even though the L\"uders reduction process generates an increase of
entropy in qualitative agreement with the second law of thermodynamics, it
cannot reproduce the quantitative time dependence of $S_{\textrm{e}}$. Previous
analyses of entropy changes during the measurement process \cite{Zeh2007} have
not taken this into account. Attaining the equality $S(\hat{\rho}) =
S_{\textrm{e}}$ was in fact the reason why von Neumann introduced his
equivalence-class definition of quantum states (for the special case of entirely
integrated subsystems) in the first place \cite{vonNeumann1955secV4}.

The standard formulation of the quantum theory of measurement is therefore
fundamentally flawed. To avoid this discrepancy with experiment it would be
necessary to restrict the domain of applicability of quantum mechanics to
microscopic subsystems (excluding measuring apparatuses), a step that Bohr
decisively rejected \cite{Bohr1928,Bohr1935,Bohr1949}. As Peres has wryly
observed \cite{Peres1995}, such a restriction would be tantamount to treating
measurement as a ``supernatural event.''

The two reduction processes also yield somewhat different predictions for the
results of measurements on isolated subsystems, although here the difference is
more subtle. The difference occurs because the L\"uders reduction process
selects one particular state in the manifold $\mathcal{M}_i^A$---generally not a
maximum-entropy state. This difference should, however, be statistically
insignificant, at least in the short term, because nearly all of the states in
$\mathcal{M}_i^A$ would lead to the same short-term dynamics of the collective
variables.

Daneri, Loinger, and Prosperi \cite{Daneri1962} used ergodic theory to justify a
reduced density operator that is similar to Eq.\ (\ref{eq:rho_hat_new}), except
that their $\rho_i^B$ is a pure state obtained by direct reduction. Such a
definition generally disagrees with the result (\ref{eq:rho_i_A_B}) of indirect
reduction and thus disagrees with experiment. These authors also did not point
out the experimental implications of their theory for the entropy of macroscopic
subsystems. Their formulation of the reduction process thus never managed to
supersede the standard formulation, and it has since fallen into half a century
of disuse.

\paragraph*{Consistency of the theoretical description.}

Although $\rho$ and $\hat{\rho}$ correspond to the same quantum state
$\eqcl{\rho} = \eqcl{\hat{\rho}}$, one may question whether the replacement of
$\rho$ with $\hat{\rho}$ could lead to any inconsistency in the theoretical
description. The key to avoiding inconsistency is that the time interval $\Delta
t$ between reductions should be chosen to satisfy
\begin{equation}
\tau_{\text{dec}} \ll \Delta t \ll \tau_{\mathcal{P}} , \label{eq:timescales}
\end{equation}
in which $\tau_{\text{dec}}$ is the maximum decoherence time for the integrated
subsystems in $A$ and $\tau_{\mathcal{P}}$ is the timescale for changes in the
projector set $\mathcal{P}$. The condition $\Delta t \gg \tau_{\text{dec}}$
ensures that different reductions do not interfere with one another, whereas
$\Delta t \ll \tau_{\mathcal{P}}$ expresses the requirement that $\mathcal{P}$
change slowly in time. The latter is a stability constraint that ensures the
reduction process $\rho \to \hat{\rho}$ can be regarded as emergent
\cite{Foreman2019a} \emph{prior} to the introduction of the equivalence class
$\eqcl{\rho} = \eqcl{\hat{\rho}}$. The satisfaction of the conditions
(\ref{eq:timescales}) is contingent upon the initial quantum state and the
definition of $\mathcal{P}$. The latter is chosen here in accordance with a
modified version \cite{Foreman2019a} of Zurek's predictability sieve
\cite{Zurek2003}, in which $\mathcal{P}$ is defined so as to generate the least
entropy during the interval $\Delta t$.

Another conceivable source of inconsistency has been discussed by d'Espagnat
\cite{dEspagnat1976}. This arises because a measuring apparatus in $A$ and its
corresponding measured subsystem in $B$ are entangled in $\rho$ but not in
$\hat{\rho}$. One could therefore conceivably perform a measurement on the
composite system (of apparatus and isolated subsystem) that would reveal this
entanglement, thus demonstrating an inconsistency in the reduction process $\rho
\to \hat{\rho}$.

The reason why this is not an actual source of inconsistency has nothing to do
with the fact that such a secondary measurement would be technologically
demanding. It is instead due to the fact that a measuring apparatus can only
function as such if it is truly integrated
\cite{Heisenberg1958_p57,PeresZurek1982,Peres1986} (i.e., if it interacts
strongly with its environment, thereby generating significant entanglement on
the timescale $\tau_{\text{dec}}$). In order for a secondary measurement to
reveal the entanglement between the primary apparatus and the isolated
subsystem, it would therefore be necessary to isolate the primary apparatus from
its environment (e.g., by cooling it down to a very low temperature, among other
things). But this changes the experimental arrangement in such a way that the
primary apparatus no longer functions as a measuring device. The secondary
measurement thus fails to demonstrate an inconsistency in the theoretical
description of the primary measurement; it merely replaces one experiment with
an entirely different experiment. This illustrates once again the importance of
the concept that ``unperformed experiments have no results''
\cite{Bohr1935,Peres1978}.

Conceptual difficulties of the type described above can often be avoided by
thinking of a measurement as something that we \emph{infer} has happened from
the correlations between subsystems in the reduced state $\hat{\rho}$, rather
than an \emph{active intervention} performed by an agent upon another subsystem.
It should be stressed here that the presumed freedom of an experimenter to
manipulate the conditions of an experiment is actually a \emph{necessary}
prerequisite for the pragmatic description of experimental physics
\cite{Bohr1935,PeresZurek1982}. The use of descriptive language involving
active agents is therefore not incorrect---however, the validity of such a
description emerges only on a timescale $\tau_{\text{fw}} \gg
\tau_{\mathcal{P}}$, when the behavior of the agent becomes so unpredictable
that the concept of ``free will'' can be invoked without fear of contradiction.
Maintaining a conceptual distinction between the emergence of outcomes
\cite{Foreman2019a} on the timescale $\Delta t$ and the emergence of free will
on the timescale $\tau_{\text{fw}}$ is therefore crucial for logical
consistency.

\paragraph*{Origins in decoherence theory.}

The mathematical form of the reduced state $\hat{\rho}$ in Eq.\
(\ref{eq:rho_hat_new}) was derived from two of the core principles of
complementarity: (1) that we obtain information about isolated subsystems
exclusively through their interactions with integrated subsystems and (2) that
unperformed experiments have no results. These principles have hitherto been
treated more or less as axioms of the Copenhagen interpretation.

However, both are corollaries of the most basic principle of 
decoherence theory---namely, that all information in quantum mechanics 
must be extracted from structures in the quantum state that are stable in time.
This principle of dynamical stability is the reason why pointer variables
are required to be robust under environmental monitoring.  As discussed previously,
it is implemented mathematically by using the predictability sieve to 
identify stable subsystems and projectors.

Principle (1) is just a special case of the dynamical stability principle,
because the integrated subsystems in principle (1) are derived from the
predictability sieve. Examples of such subsystems include measuring instruments
whose pointers are described by quasiclassical collective variables. Isolated
subsystems, by contrast, are fragile and have no stable properties that can be
defined independently of the integrated subsystems with which they interact.

Principle (2) then follows immediately, because the dynamically stable pointer
variables are not arbitrary. Subsystems and basis states can indeed be chosen
arbitrarily, but essentially only one such configuration yields dynamically
stable experimental information. A given quantum state cannot be described in
terms of different experimental arrangements leading to different (i.e.,
complementary) types of experimental information. All experimental arrangements
other than the dynamically stable one must be regarded as unperformed.

\paragraph*{Common ground.}

As shown above, the principle of dynamical stability in decoherence theory leads
to several key elements of Bohr's principle of complementarity, which in turn
have definite implications for the mathematical structure of the quantum state.
The expression of these results in mathematical form clarifies the meaning of
complementarity and makes it easier for everyone to ``get the hang of it.''
However, this mode of derivation also shows that the significance of these
results extends beyond the philosophy of the Copenhagen interpretation. The
extraction of dynamically stable experimental information is an operational
issue that must be dealt with at some point by every interpretation of quantum
mechanics. This is highlighted by the fact that the modified definition of the
quantum state proposed here has unambiguous experimental implications for the
entropy of macroscopic subsystems. This establishes an area of common ground
that may help to bring different workers in the field of quantum foundations
closer together.

% Create the reference section using BibTeX:
% \bibliographystyle{abbrv}

%\bibliography{quantum}

%apsrev4-2.bst 2019-01-14 (MD) hand-edited version of apsrev4-1.bst
%Control: key (0)
%Control: author (8) initials jnrlst
%Control: editor formatted (1) identically to author
%Control: production of article title (0) allowed
%Control: page (0) single
%Control: year (1) truncated
%Control: production of eprint (0) enabled
%

\end{document}